# Study and characterization of GaN MOS capacitors: planar versus trench topographies

K. Mukherjee[1], C. De Santi[2,*], S. You[3], K. Geens[3], M. Borga[3], S. Decoutere[3], B. Bakeroot[4], P. Diehle[5], F. Altmann[5], G. Meneghesso[2], E. Zanoni[2], M. Meneghini[2]

[1]formerly with the University of Padova, Department of Information Engineering, Via Gradenigo 6/B, 35131 Padova, Italy.
[2]University of Padova, Department of Information Engineering, Via Gradenigo 6/B, 35131 Padova, Italy.
[3]imec, Kapeldreef 75, Heverlee, Belgium.
[4]CMST, imec & Ghent University, Technologiepark 126, 9052 Ghent, Belgium
[5]Fraunhofer Institute for Microstructure of Materials and Systems IMWS, Walter-Huelse-Strasse 1, 06120 Halle, Germany.
* carlo.desanti@unipd.it

*Abstract*: Developing high quality GaN/dielectric interfaces is a fundamental step for manufacturing GaN vertical power transistors. In this paper, we quantitatively investigate the effect of planar etching treatment and trench formation on the performance of GaN-based MOS (metal oxide semiconductor) stacks. The results demonstrate that (i) blanket etching the GaN surface does not degrade the robustness of the deposited dielectric layer; (ii) the addition of the trench etch, while improving reproducibility, results in a decrease of breakdown performance compared to the planar structures. (iii) for the trench structures, the voltage for a 10 years lifetime is still above 20 V, indicating a good robustness. (iv) To review the trapping performance across the metal-dielectric-GaN stack, forward-reverse capacitance-voltage measurements with and without stress and photo-assistance are performed. Overall, as-grown planar capacitors devoid of prior etching steps show lowest trapping, while trench capacitors have higher interface trapping, and bulk trapping comparable to the blanket etched capacitors. (v) The nanostructure of the GaN/dielectric interface was characterized by high resolution scanning transmission electron microscopy (HR-STEM). An increased roughness of 2-3 monolayers at the GaN surface was observed after blanket etching, which was correlated to the higher density of interface traps. The results presented in this paper give fundamental insight on how the etch and trench processing affects the trapping and robustness of trench-gate GaN-MOSFETs, and provide guidance for the optimization of device performance.

The formation of the gate module is of critical importance to the functioning of metal-oxide-semiconductor field effect transistors (MOSFETs), which are subject of intense research[1,2]. In GaN-based vertical MOSFETs, the output current is maximized through a large number of densely packed trenches, each of which comprises the gate electrode and the gate dielectric interfacing with GaN.

Several processing features are necessary for trench fabrication, including (i) a controllable dry etch process to reduce surface roughness around the trench, and (ii) rounded corners to minimize field-crowding, followed by (iii) a cleaning process to ensure the minimization of impurities and residuals dependent on the process[3–8]. Similar steps may be necessary also for MESA fabrication in vertical diode structures[9].

Each of these steps may degrade the quality of the surface, thus potentially lowering the stability (due to trapping effects) and the reliability of the devices. Despite the importance of this topic for future vertical GaN technologies, the understanding of the impact of process steps on dielectric/interface properties is still limited.

The goal of this paper is to improve the understanding of the relation between trench process parameters and trapping phenomena in GaN-based MOSFETs. To this aim, we present an extensive and systematic analysis of the impact of blanket etch, selective etch and trench formation on the robustness and stability of GaN-based MOS capacitors.

We demonstrate that a) a blanket etch prior to gate deposition does not alter the robustness of the devices, but results in an increase in the trap density (+ 44 % bulk and + 30 % interface trapping), compared to the as-grown (unetched) sample. In addition, b) the formation of the trench may result in a decrease in breakdown voltage (-10 V), due to the higher electric field present at the trench corners. Finally, c) the formation of the trench results in an increased interface trapping (+ 50 % bulk and 45 % interface trapping with respect to the as-grown case). Photo-assisted capacitance voltage measurements and high resolution transmission electron microscopy were carried out to gain insight on the observed trapping processes.

We investigated the reliability and trapping performance of MOS capacitor structures. To fabricate these capacitors a $n^+$-GaN layer with $6 \times 10^{18}$ Si atoms/cm$^3$ is grown on top of an uid-GaN layer and some stress compensation layers on a Si substrate, by MOCVD. All the structures have a bilayer dielectric stack composed of an interfacial 2.5 nm layer of $Al_2O_3$, and a bulk $SiO_2$ layer of 50 nm, deposited on the top $n^+$ GaN layer. The bilayer configuration is advantageous in trench MOSFETs as reported in previous works[10,11], as the ALD-deposited thin dielectric ensures a high-quality interface with GaN, while the thick $SiO_2$ improves the gate dielectric stack robustness. While emulating the behavior of unit gate cells of a trench power MOSFET, the larger area (= $7.25 \times 10^{-4}$ cm$^2$) of the test capacitors allows for improved resolution and diagnostic validity of the measured data.

The three tested wafers differ in the semiconductor etching type. The "reference" wafer has planar capacitors on as-grown (i.e. devoid of any prior etching steps) gallium nitride, to reflect the properties of the dielectric stack and of the interface when the dielectric is deposited on as-grown $n^+$-GaN. The "blanket" wafer has planar capacitors fabricated on a blanket etched $n^+$-GaN surface, to study the impact of the bare etching process and to reproduce the conditions expected along the bottom of a trench. Finally, the "trench" wafer has trench capacitors with a selective deep trench etch to analyze the effects of the complete trench geometry, including the trench sidewalls and corners. The trapezoidal shape of the gate trench was found to be beneficial[12] due to an optimized electric field profile, in addition to providing better homogeneity for the top metal coverage. For the wafer with a blanket GaN etch and the wafer with a patterned trench etch, an identical dry etch step was used for GaN removal, in a biased ICP (Inductive Coupled Plasma) etch chamber. First, a bulk GaN removal step was implemented using a $Cl_2$/Ar chemistry. Secondly, Atomic Layer Etch (ALE) processing steps are implemented as a soft landing step[11]. Before the dielectric deposition is done, wet cleaning steps are implemented to clean up organic residues. On top of the gate dielectric a TiN/Ti/Al containing stack is used as metal electrode. Afterwards a N-implantation is used to isolate the $n^+$-GaN in the bond pad areas of the capacitor structures. Ohmic contacts to the $n^+$-GaN are realized by recess etching, cleaning, Ti/TiN/Al based metallization and low-temperature ohmic anneal[13]. PECVD $SiO_2$ is used as intermetal dielectric and the metal stacks are finished with a 4 µm-thick Al stack. The schematic cross-section of the capacitors is shown in Fig. 1 (a).

Initially, we compared the breakdown behavior of the three wafers: 9 capacitors from each wafer were subjected to I-V sweeps until hard breakdown, and the statistical distribution of the breakdown voltage ($V_{BR}$) is compared in Fig. 1 (b).

First, we notice that the "blanket" wafer shows only a small worsening compared to the reference wafer, indicating that a planar blanket etch does not significantly impact on device reliability. On the other hand, the "trench" wafer has a 20 % lower $V_{BR}$, indicating that the 3D trench structure results in a reduced reliability. We notice that the intra-wafer dispersion in $V_{BR}$ is lower for the trench wafer.

To obtain an estimation of expected lifetimes, constant voltage stress tests on 9 devices from each wafer were performed until breakdown. The average $V_{BR}$ for each voltage and wafer is summarized in Fig. 1 (c), along with the lifetime extrapolation. As can be noticed, the results indicate that the voltage for 10-year lifetime is lower for the "trench" wafer (25 V), compared to the "reference" wafer ($\approx$ 40 V, Wafer A). To understand these results, TCAD simulations were carried out with Sentaurus, the device simulation tool by Synopsys. The results (presented in Fig. 2 for a representative voltage of 30 V) demonstrate a stronger electric field crowding at the trench corners for the "trench" wafer, with an overall electric field peak significantly higher than in planar capacitors: this results in a lower breakdown voltage compared to the planar capacitors of the "reference" and "blanket" wafers. Additional details on the simulation framework can be found in the supplementary material.

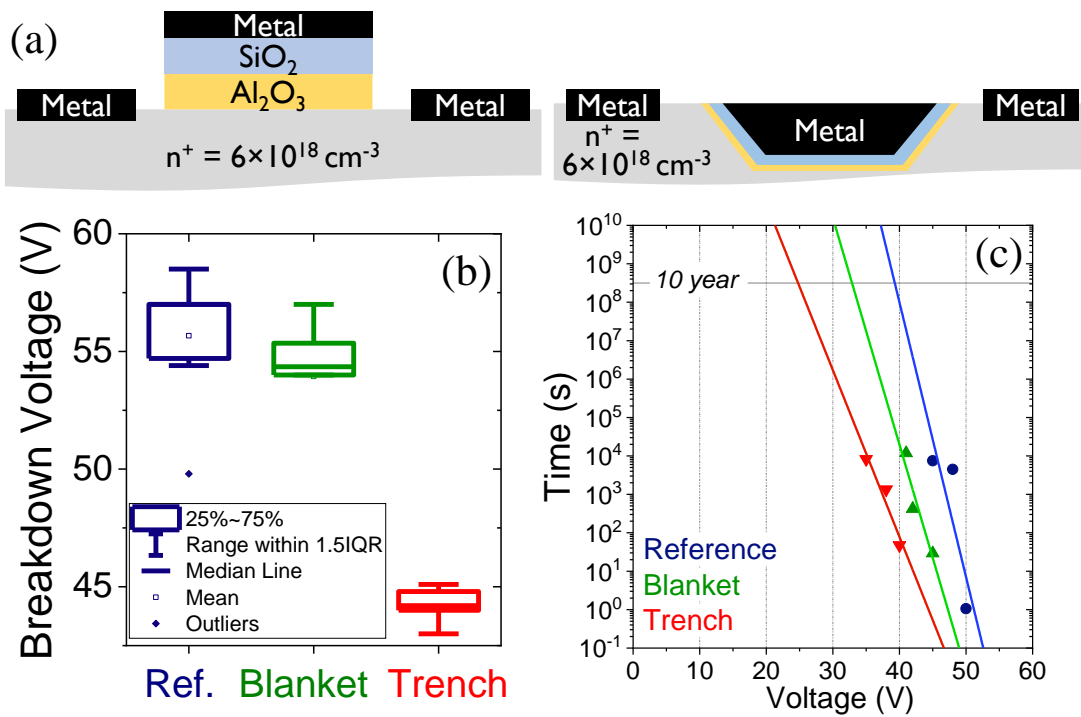

Fig. 1. (a) Schematic structure of the planar and trench capacitors. (b) Average breakdown voltage from I-V sweeps and (c) estimated lifetime from constant voltage stress measurements, for the different wafers. Blanket etch implies only a small reduction in breakdown voltage, while trench formation results in a significant decrease. It is worth noticing that voltage for 10 years lifetime will be higher than 20 V for all analyzed samples.

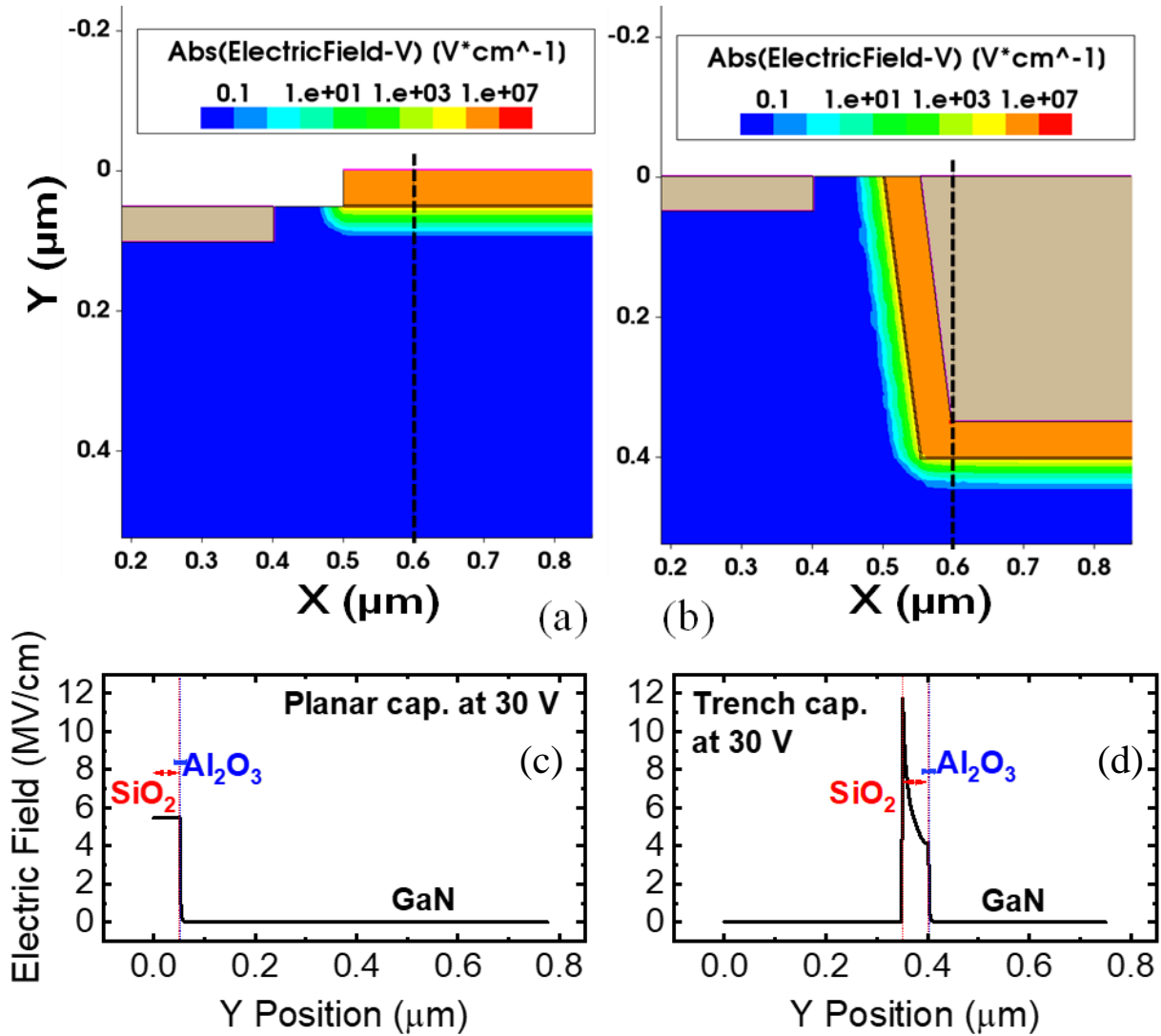

Fig. 2. Simulations of the electric field distribution for V = 30 V across (a) planar and (b) trench capacitor structures. (c) and (d) are electric field profiles along the cut-lines in (a) and (b) respectively. The results confirm that the formation of the trench results in a higher field across the insulator, which leads to a lower breakdown voltage for the trench wafer.

To compare the trapping in the bulk and at the interface of the deposited dielectric, a set of C-V measurements were performed. First, C-V hysteresis was measured, from 0 V to increasing stop voltages, to calculate the effective traps density ($N_{eff}$) profile for each wafer, shown in Fig. 3 (a). $N_{eff}$ estimated in this way was found to be comparable between the three wafers. However, in simple forward and backward sweeps, the trapping time is relatively short: for this reason, to compare the wafers in a worst-case scenario, the testing conditions were aggravated, by forcing a trapping time at a given positive voltage. Specifically, bi-directional CV measurements from deep depletion at -25 V to accumulation at 25 V were repeated for each wafer, with the addition of a stress period of 120 s at $V_G$ = 25 V between the forward and backward CV sweeps.

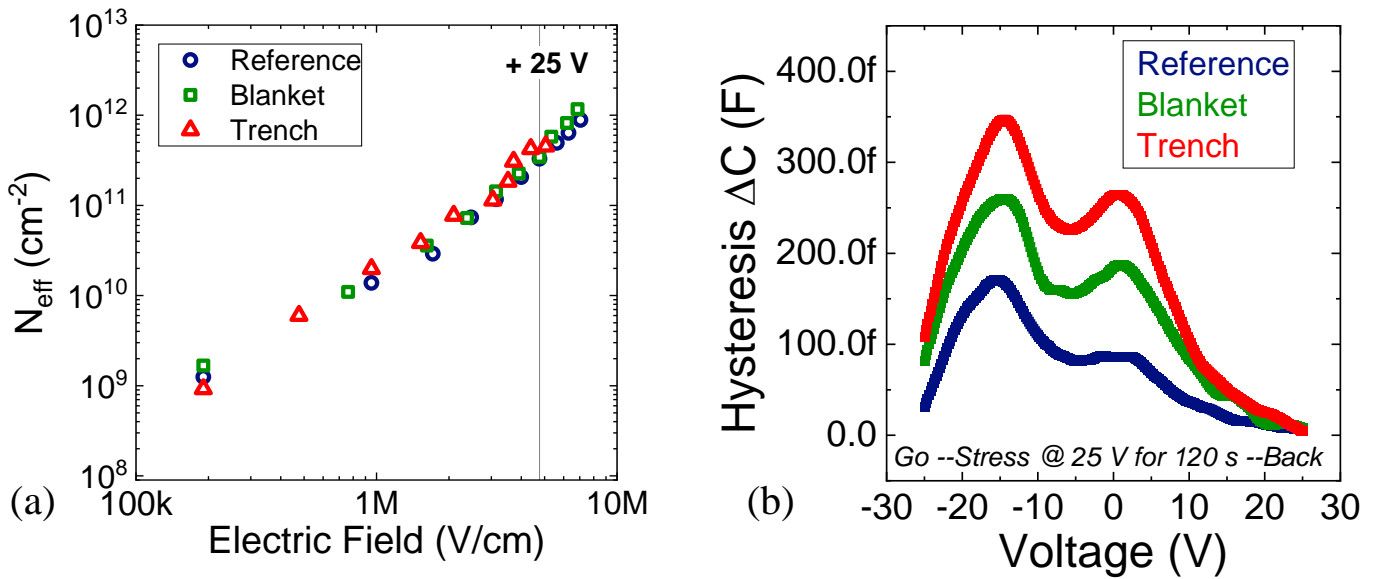

Fig. 3. (a) Comparison of effective trapped charge from bi-directional CV measurements, for the different wafers. The fast measurement protocol does not identify differences among the wafers, so a more detailed analysis is proposed. Specifically, (b) reports the capacitance hysteresis measured after the introduction of a trapping time between the forward and backward C-V sweep. This approach results in measurable charge trapping. Blanket etch and trench formation result in a larger interface/oxide trapping, compared to the reference process

Fig. 3 (b) shows the comparative hysteresis magnitudes between the wafers. As can be noticed, every etching process results in increased trapping; in addition, the trench etched capacitors display the largest hysteresis, possibly an effect of the different geometry and non-uniform electric field (which may favor a stronger trapping) or of the presence of different crystal planes involved, leading to a different amount of damage in the planar and diagonal region of the trench. In all cases, the maximum hysteresis between the forward-backward curves is observed around -15 V, and is associated to a) a higher trapping at the $Al_2O_3$/GaN interface[14,15], since a change in the slope of the backward curve is detected, and b) a larger oxide trapping.

To quantitatively evaluate the amount of oxide and interface traps in the three wafers, photo-assisted CV analysis was carried out. This allows to (i) accurately extract the total amount of trapped charge by starting from a completely de-trapped device state, and (ii) to reliably separate the interface ($D_{it}$) and stored bulk oxide trap ($Q_{ox}$) contributions.

The principle of the photo-assisted CV experiment[15] is illustrated in Fig. 4 (a). (1) First, UV (wavelength = 365 nm) illumination (50 s of LED illumination followed by 500 s of no illumination) is applied, to induce photoionization of energetically deep interface/bulk trap states; then, the (2) forward C-V sweep is measured from -25 V to +25 V. Such a "forward" characteristic represents the ideal, completely de-trapped device response.

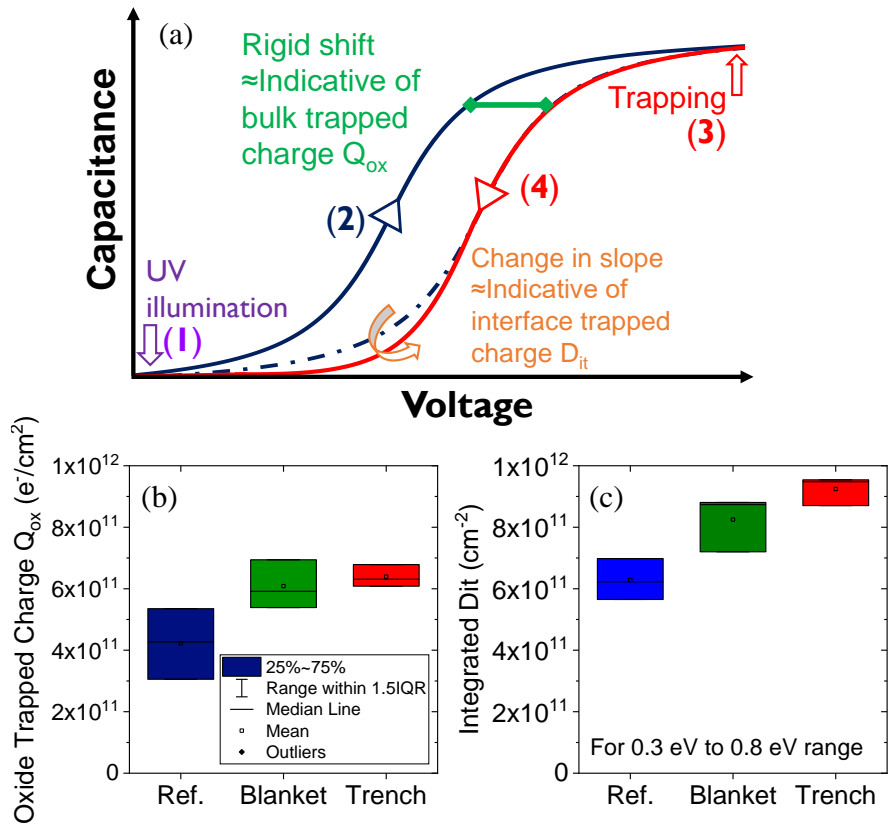

Fig. 4. (a) Principle of the photo-assisted C-V procedure. Comparison of (b) oxide trapped charge and (c) integrated $D_{it}$ for the three wafers considered within this paper. Blanket etch results in a moderate increase in bulk and interface states; trench processing further increases interface state density

At this point, the traps are filled by (3) stressing for two minutes at high positive voltage (25 V). Finally, the (4) backward sweep is performed from +25 V to -25 V. Representative capacitance-voltage curves after illumination and after filling in accumulation can be found in the supplementary material.

By separately evaluating $Q_{ox}$ (extracted from the absolute rigid shift between the forward and backward curves where the $D_{it}$ impact is minimum) and $D_{it}$ (extracted from the change in the slope between forward-backward curves), the total trapped bulk oxide charge and interface charge can be estimated, as compared for all the wafers in Fig. 4 (b) and (c).

It is clear that both blanket etching and trench formation result in the worsening in bulk trapping ($Q_{ox}$), with comparable magnitude, compared to the reference wafer (Fig. 4 (b)). Compared to the reference case, $Q_{ox}$ is increased by 44 % and 51 % for the blanket and the trench wafer, respectively. With regard to interface traps ($D_{it}$), the extracted energy profile for the traps was found to lie between 0.3 to 0.8 eV; interface trap density was found to rise steadily from the reference to the "trench" wafer (Fig. 4 (c)).

Specifically, the blanket wafer displays a $D_{it}$ 31 % higher than the reference wafer, while the corresponding increase in trench wafer is ≈ 45 %. It is worth noting that the total amount of trapped charge in Fig. 4 is larger than the extracted values in Fig. 3, which confirms that the photo-assisted CV method provides a more complete filling of defects, and a more accurate extraction of the total (traps empty to traps filled) trapped charge compared to a simple forward/backward CV sweep. The fact that a simple bi-directional CV sweep is not sufficient for a complete characterization of trapping phenomena in GaN-based MOS stacks can be ascribed to the relatively long trapping times of the analyzed samples, which are in the order of seconds (see similar previous reports on complete transistors[16]).

To understand the impact of the blanket etch process on the electrical behavior of the interface, microstructural analysis of the dielectric was carried out for the reference and blanket planar capacitors. Cross sectional, electron transparent films located at the middle of the capacitors were prepared by large area Ar-ion milling to guarantee a Ga-free sample preparation providing best conditions for atomic resolved TEM imaging. EDXS (energy dispersive X-ray spectroscopy) and HR-STEM (high resolution – scanning transmission electron microscopy) investigation were performed on a probe-corrected 200 kV Hitachi HF 5000 TEM equipped with a double EDX detector. In Fig. 5 (a), the two amorphous dielectric layers of the reference planar capacitor can be seen. The corresponding EDXS analysis of these region of interest shows a homogenous 2.5 nm thin $Al_2O_3$ layer followed by a 50 nm thick $SiO_2$ layer (see Fig. 5 (b-e)). No difference in the thickness of the individual layers or chemical composition were observed for the reference and blanket etched planar capacitors. Additional high magnification EDXS investigation of the $GaN/Al_2O_3$ interface showed no substantial deviations between both samples (not shown). DF STEM images (not shown), with a 100 nm scale, were obtained to compare the semiconductor surface with and without blanket GaN etch. The surface roughness is comparable, and more details can be found in ref. [17].

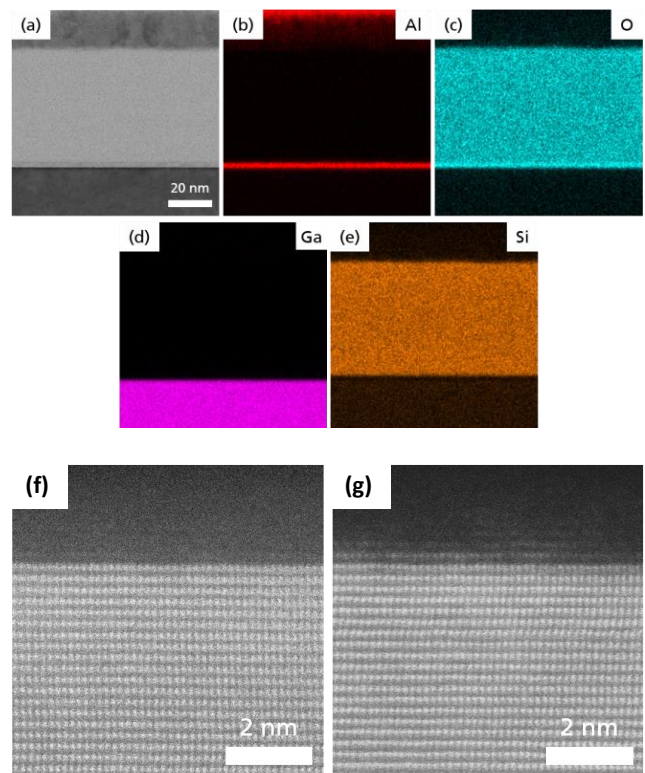

Fig. 5. (a) BF-STEM image and (b-e) corresponding elemental EDXS maps showing the dielectric layers of the reference planar capacitor. The 2.5 nm thin $Al_2O_3$ layer is homogenous in thickness and forms an interface to GaN. The $Al_2O_3$ layer is followed by a 50 nm thick $SiO_2$ layer. HAADF-HR-STEM images of GaN – $Al_2O_3$ interface for (f) reference and (g) blanket etched planar capacitors. Blanket etching results in an increased interface roughness, that can explain the increased charge trapping revealed by electrical measurements.

Then, the $GaN/Al_2O_3$ interfaces of both planar capacitors were characterized by means of HR-STEM (see Fig. 5 (f-g)). The reference capacitor has a sharp $GaN/Al_2O_3$ interface with an atomic flat GaN surface. In contrast, the GaN surface of the blanket planar capacitor has a roughness of 2-3 monolayers. The increased roughness could have enhanced the presence of interface defects on the etched capacitors, and/or favored carrier leakage from the semiconductor to the insulator, thus resulting in stronger trapping phenomena.

In summary, we evaluated the impact of specific processing steps (blanket etch, selective etch, trench formation) on the stability and robustness of MOS capacitors: this aspect is fundamental for the development of vertical GaN MOSFETs.

The results of our analysis indicated the following. (i) The blanket etching of the GaN surface results in increased trapping (by an average of 35 %) compared to oxide deposition on as-grown (un-etched) GaN films. However, the breakdown robustness is not compromised. (ii) Introducing a trapezoidal-shaped trench etch of controlled depth reduces the breakdown capability by an average of 10 V due to worsened field crowding effects, and results in increased trapping. While the bulk trapping effects remain comparable to that of planar capacitors on blanket etched GaN, the interface trapping at the oxide/semiconductor interface and/or border traps is strengthened further (+ 10 % compared to blanket etched case, and + 45 % compared to the as-grown case), possibly due to the combined effect of the blanket etch and of the sidewall formation. (iii) The microstructural quality of the insulator and interfaces was analyzed by means of HR-STEM and EDXS. No impact of the blanket etching on the chemical composition of the dielectric layers for the as-grown and the blanket etched planar capacitors were observed. The blanket etching induced a degradation of the GaN surface resulting in an increased roughness of 2-3 monolayers compared to an atomic flat as-grown GaN surface, that can explain the stronger charge trapping phenomena observed after etching. The results described within this paper can be used as guidelines for the optimization of MOS stacks to be used in vertical GaN transistors.

SUPPLEMENTARY MATERIAL

See supplementary material for details on the simulation framework and for representative photo-assisted C-V measurements.

ACKNOWLEDGMENTS

This work was carried out within the UltimateGaN project, that has received funding from the ECSEL Joint Undertaking (JU) under grant agreement No 826392. The JU receives support from the European Union's Horizon 2020 research and innovation programme and Austria, Belgium, Germany, Italy, Slovakia, Spain, Sweden, Norway, Switzerland. The UltimateGaN project is co-funded by the Ministry of Education, Universities and Research in Italy. The authors would like to thank A. Böbenroth for TEM sample preparation.

AUTHOR DECLARATIONS

The authors have no conflict to disclose. The data that support the findings of this study are available from the corresponding author upon reasonable request.